\documentclass[]{spie}  

\usepackage{amsmath,amsfonts,amssymb}
\usepackage{graphicx}
\usepackage{xcolor}
\usepackage[colorlinks=true, allcolors=blue]{hyperref}
\usepackage{subcaption}
\usepackage{placeins}
\usepackage{booktabs}
\usepackage{cite}

\usepackage[printonlyused]{acronym}

\definecolor{navy}{HTML}{000F8A}
\definecolor{blueviolet}{HTML}{106EB8}

\hypersetup{
    colorlinks=true,
    linkcolor=navy,
    filecolor=navy,      
    urlcolor=navy,
    pdfpagemode=FullScreen,
}

\newcommand{\sref}[2]{\hyperref[#2]{#1\ref*{#2}}}

\newlength{\cellwidth}
\title{A New AI/ML Based Image Reconstruction Technique for Hard X-ray Coded Aperture Imagers}

\author[a]{Shayla Sharmin}

\author[a]{Branden T. Allen}

\affil[a]{Space Science and Engineering Initiative (SSEI), College of Engineering (CoE), Institute for Astronomy (IfA), University of Hawai'i at M\={a}noa, HI 97620, USA}

\authorinfo{Send correspondence to Shayla Sharmin\\ E-mail: shay28@hawaii.edu}

\begin{document} 
\maketitle

\begin{abstract}
Here we demonstrate a new AI/ML based image reconstruction method for coded-aperture instruments that provides a pathway for embedded real-time use on-board spaceflight missions. Using the Swift Burst Alert Telescope (BAT) for our initial simulated demonstration, we used our new technique to replace the standard Fast Fourier Transform (FFT) based image reconstruction pipeline. In our alternative approach, we have created a neural network that maps detector counts directly to source positions using simulated responses from selected Swift Burst Alert Telescope (BAT) observations. The network combines convolutional feature extraction with attention and is trained at around 900 source counts. Under zero background noise, it achieves over 93\% exact-pixel localization accuracy with 400 detected source photons. The model maintains at least 90\% exact-pixel localization accuracy with up to 8{,}500 added background counts for sources with 900 detected photons. Within the evaluated region of BAT's field of view, these results show that source position can be recovered from sparse coded-aperture measurements in a wide-field instrument. A key feature of this technique is that explicit inclusion of the mask pattern is not required. This allows us to train the algorithm to "see" using data accumulated from an imaging system. These new non-traditional imaging techniques show substantial promise in enabling efficient imaging using non-planar geometries, which will enable new classes of hard X/$\gamma$-ray imaging systems and simultaneously improve imaging capability for currently operating instruments.
\end{abstract}


\section{INTRODUCTION}
\label{sec:intro}  
Imaging X/$\gamma$-ray sources below the Compton regime (below approximately 300~keV), across a wide field of view greater than 100~deg.$^2$, and at moderate angular resolution (approx. 1$^\prime$-30$^\prime$), is frequently accomplished through coded-aperture imaging \cite{Goldwurm2024}. Conventional focusing optics typically offer limited fields of view on the order of 0.25~deg.$^2$, and focusing becomes increasingly impractical at higher photon energies, in particular above 10~keV), where complex grazing-incidence optical architectures and assemblies with long focal lengths on the order of 10~m \cite{harrison2013nustar}. Coded-aperture imaging addresses both constraints by using a patterned mask to encode the direction of incoming photons across a wide field of view into its detector measurements. The information available for reconstruction in such systems therefore depends on both the mask pattern and its geometric relationship to the detector, as well as other characteristics such as the relative mask and detector pixel sizes and the presence of support structure and gaps in sensitive area, mask and shielding thickness. \cite{skinner2008sensitivity}.

The Burst Alert Telescope (BAT) aboard the Neil Gehrels Swift Observatory shown in \sref{figure }{fig:bat} is a wide-field coded-aperture instrument sensitive between 15–150~keV. Its well-characterized system response, software, and observational archive provide a suitable platform for developing and testing new reconstruction techniques. Source recovery in BAT is conventionally performed by cross-correlating the detector image with the mask pattern, which is a linear reconstruction operation that is computed efficiently via the fast Fourier transform (FFT) \cite{barthelmy2005bat}. However, this simple FFT-based formulation relies on an idealized geometry in which the mask and detector are flat and parallel, a configuration that preserves the shift invariance property of convolution for distant sources. This imposes structural constraints on instrument design, particularly for on-board prompt detection and localization of transient sources. Complex detector geometries with non-planar arrangements of the mask and detector can be optimized to increase photon collection, widen the accessible field of view, and preserve shadow information that would otherwise be lost when sources are imaged near the boundary of a flat detector \cite{zelakiewicz2011soris}. Once the geometry deviates from a parallel planar configuration, the mapping from sky to detector is generally no longer shift invariant, and the same computationally efficient FFT-based cross-correlation cannot be applied directly. The choice of reconstruction algorithm therefore constrains both the achievable localization accuracy and the instrument geometries practical for real-time imaging.

\begin{figure}[htbp]
    \centering
    \includegraphics[width=0.5\linewidth]{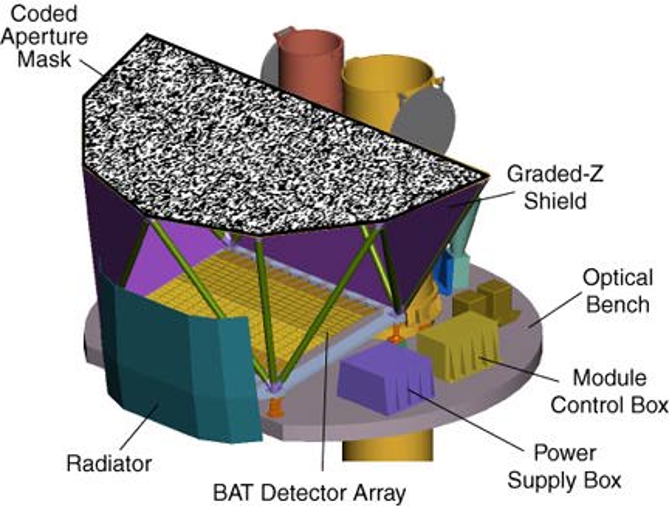}
    \vspace{0.5cm}
    \caption{The Burst Alert Telescope (BAT)\cite{nasa_gsfc_swift_description}}
    \label{fig:bat}
\end{figure}
Deep learning based methods have been investigated as alternatives to conventional coded-aperture reconstruction for various types of systems. Zhang et al. \cite{zhangCNN} trained a small CNN-based model with a fully connected sigmoid output to map a \(17~\times~17\) CZT detector shadowgram to a \(17~\times~17\) source-location map. Although the model generally produced higher-CNR reconstructions than correlation decoding, it sometimes predicted high-intensity sources at incorrect locations. Subsequent work used MLP with the same compact CZT detector but a random coded mask \cite{Wang}. Experimental measurements with unshielded background showed higher CNR and localization accuracy than correlation decoding. However, the reported accuracies were 75\%, 80\% and 88\% for mask opening fractions of 50\%, 70\%, and 90\%, respectively. Both models rely on fully connected output mapping, which is an impractical approach for wide-field systems like BAT as it requires a substantially larger reconstruction grid. Mei{\ss}ner et al. \cite{CED} used a fully convolutional encoder-decoder with 16.4 million parameters that reaches 512 channels at bottleneck, to reconstruct \(256\times256\) output image. Skip-connections were not utilized as there is no similarity between input and output plane. It is worth noting their proposed 16.4M parameter model required approximately \(0.29\)~s per reconstruction on a CPU, compared with \(13\)~s for 25 iterations of convolutional MLEM. Training used photon levels between \(10^3\)-\(10^6\) with Poisson fluctuations and no additional background noise. Evaluation focused on the CNR of reconstructed medical phantom images rather than accurate source localization. Overall these studies demonstrate the potential of neural networks in mitigating reconstruction noise and artifacts that limit conventional decoding in coded aperture systems. However, their applicability to localization in wide-field detectors under sparse source-photon conditions remains largely unexplored.

In this work, we propose a convolutional-transformer network to map detector inputs directly to reconstructed sky output. Without requiring an explicit analytical inverse of the system response, the model learns an end-to-end mapping from paired input-output data. It offers a potential route toward a new class of hard X-ray and $\gamma$-ray imaging systems. We demonstrate the approach using the Swift BAT software to simulate detector responses for selected observations. We show that the model generalizes to source counts below its training level, maintains robustness across a wide range of background levels, and degrades gracefully under more challenging detector conditions. These results on a known planar system provide a foundation for extending this approach to systems with non-traditional geometries, as well as a potential pathway to improve imaging capabilities for existing instruments through learned models that implicitly capture instrument features that are not or cannot be included in existing image reconstruction pipelines.

\section{Methods}\label{sec:methods}
\subsection{Dataset Generation for Supervised Learning}

Input-output pairs for supervised training were generated using the official BAT instrument response \cite{batSW}. Sources were placed on a regular grid of sky positions centered on the Crab Nebula ($\mathrm{RA} = 83.6^{\circ}$, $\mathrm{Dec} = 22.0^{\circ}$), spanning $\pm 20^{\circ}$ in each coordinate at a step of $0.5^{\circ}$, yielding $6{,}561$ unique sky positions. The sky image resolution was set to $128 \times 128$ pixels. For each source position, the corresponding clean detector map was generated by computing a ray-traced shadow pattern for a source at that location using \texttt{batmaskwtevt}. The resulting map was passed through \texttt{batbinevt} to extract the detector plane image (DPI), which was then scaled by a flux factor $f\,/\,(0.16 \times 0.8)$ to set the expected source count level on the detector. The observed measurement was subsequently drawn by Poisson sampling. Training detector samples were generated at the $900$-count level, with $10$ independent observations per source location and flux level.
\begin{figure}
    \centering
    \includegraphics[width=\linewidth]{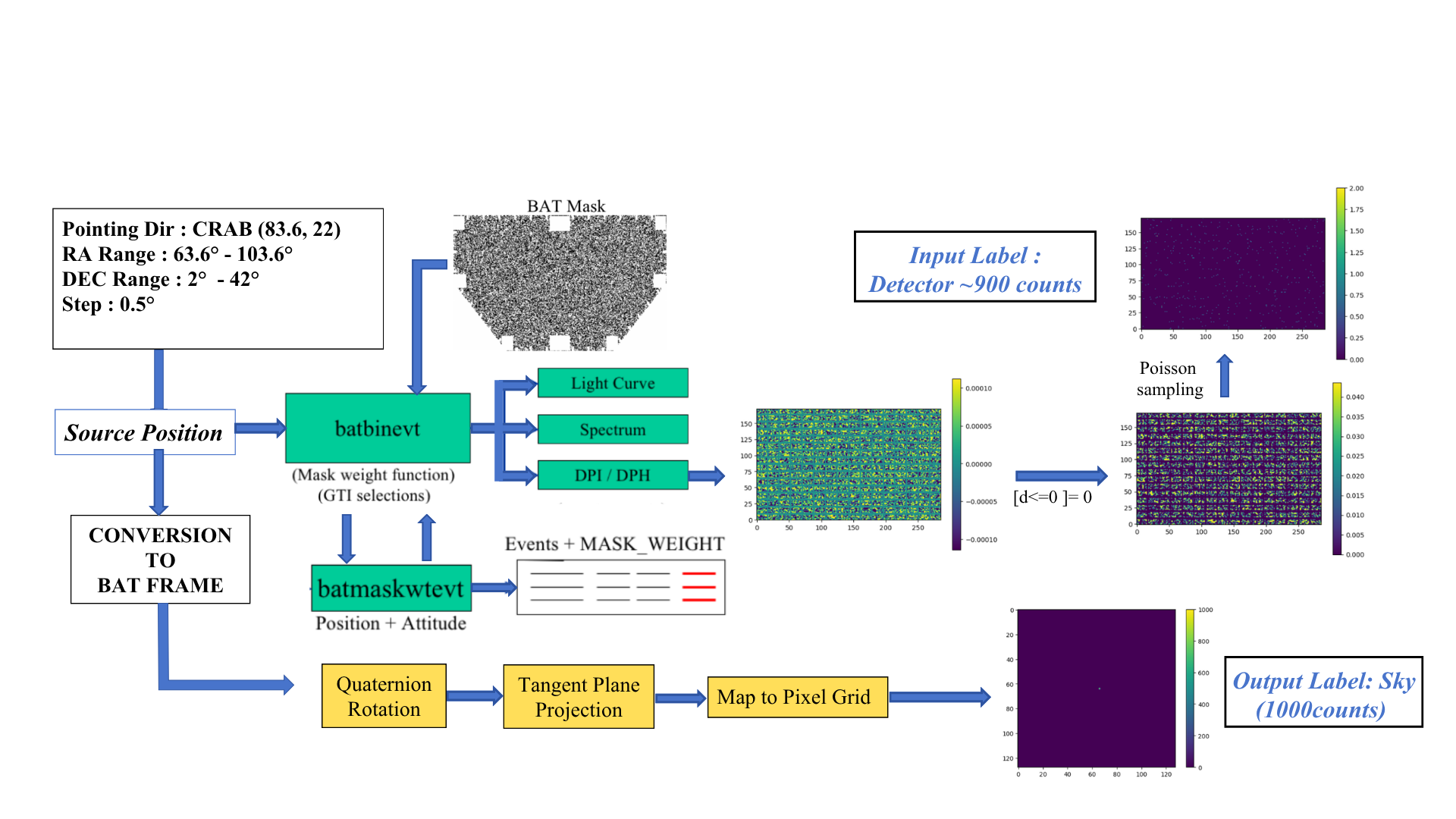}
    \caption{Dataset Generation Workflow}
\end{figure}

The corresponding target was a clean sky image in which an event of $1{,}000$ counts was placed at the true source location. Sky pixel coordinates were computed from the source's celestial coordinates as follows. A fixed quaternion rotation, derived from the boresight pointing, transformed the celestial direction into the BAT instrument frame. The rotated direction was then projected onto the sky plane via a tangent-plane projection and mapped to a pixel on the $128~\times~128$ output grid. This procedure yielded $65{,}610$ initial training samples containing no background noise. A validation set of $6{,}561$ samples was constructed with one sample per sky location to ensure a balanced distribution across the field of view. Low-count test sets were generated analogously by scaling the detector maps to count levels in the range $400$--$900$. To train the model under realistic background conditions, each sample in the ideal training and validation sets was augmented by adding uniformly scattered background counts at the following discrete levels: $100$, $300$, $500$, $700$, $900$, $1{,}000$, $2{,}000$, $3{,}000$, $4{,}000$, $5{,}000$, $6{,}000$, $7{,}000$, $8{,}000$, $9{,}000$, and $10{,}000$ counts, producing an augmented dataset of approximately one million samples. The ideal $900$-count test set was corrupted with background at levels different from those used during the training, in order to evaluate generalization to unseen background conditions.

\subsection{Model Architecture}

To demonstrate that a network can learn the system end to end from data, the model was trained without embedding any analytical reconstruction operator or explicit system characteristic. Because the encoded information regarding the source position is distributed across the entire detector plane rather than confined to a localized region, the model can benefit from capturing long-range dependencies. Applying a transformer directly to the full resolution detector image would be computationally infeasible for resource constrained flight systems. Conversely, convolutional networks with small kernels may have difficulty capturing detector-wide relationships. Although a deep convolutional encoder-decoder network with large early kernels can be used \cite{CED}, it can be insufficient under the low photon conditions encountered in high-energy astronomy. We therefore propose a hybrid architecture that combines multiscale feature extraction of cnn with global self-attention. 
The encoder starts with several $7\times7$ residual convolutional blocks processing the detector at full resolution while the number of channels remains low at 32. Three stages of downsampling reduce the spatial dimensions to $22\times36$ while increasing the number of feature channels to 256. The detailed architecture of the network is given in \sref{figure }{fig:model}. This brings the feature map to a resolution at which attention becomes possible without aggressive information loss or inflated compute cost. The feature map is flattened into a sequence of only 792 spatial tokens with an embedding dimension of 256. Learned positional embeddings are added to retain spatial ordering of the encoded features. The sequence then passes through three pre-normalized transformer blocks, each comprising eight-head self-attention, a multilayer perceptron, and residual connections. 
\begin{figure}[htbp]
    \centering
    \includegraphics[width=\linewidth]{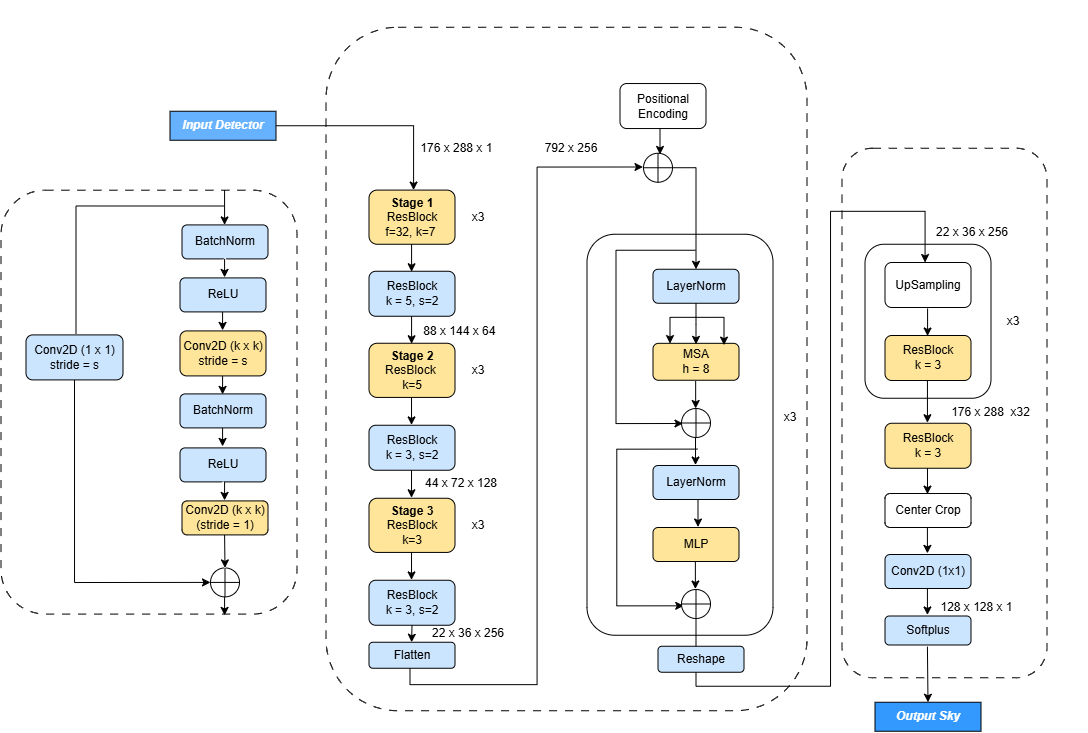}
    \caption{Network Architecture}
    \label{fig:model}
\end{figure}
The transformer output is reshaped to $22\times36\times256$ and decoded through three residual upsampling stages using $3\times3$ kernels. The tensor is then center-cropped to match the $128\times128$ target sky image dimensions, and a final $1\times1$ convolution produces the single channel intensity map. The final layer applies a softplus activation, which constrains its values to be non-negative. The model has approximately $6.2$ million parameters and produces each reconstruction in a single forward pass of around 0.25 s on a CPU during inference.

\subsection{Training Procedure}

We first train the model under ideal detector conditions using $65{,}610$ samples for $50$ epochs, with a batch size of $32$, on a NVIDIA A30 GPU. Training uses AdamW with a peak learning rate of $10^{-3}$ and a weight decay of $10^{-2}$. Following five epochs of linear warmup, the learning rate decreases through cosine annealing to $0.01$ times its peak value. Gradients are clipped at $1.0$, and the random seed is fixed at $42$. The objective used to train the network is to minimize the Poisson negative log-likelihood over all output pixels:
\begin{equation}
    \mathcal{L}_{\mathrm{Poisson}} = \sum_{i=1}^{N} \left[ \lambda_i - z_i \log \lambda_i \right],
    \label{eq:poisson}
\end{equation}
where $\lambda_i$ is the predicted intensity at pixel $i$, $z_i$ is the corresponding target value, and $N$ is the number of output pixels. To ensure that $\lambda_i$ remains positive, we apply a softplus activation to the final-layer pre-activation $a_i$, giving $\lambda_i=\log(1+e^{a_i})$. Unlike ReLU, softplus retains a nonzero gradient everywhere. Detector inputs were normalized by dividing by $5$, the maximum pixel value in the training set, and target source intensities were divided by $20$. The same scaling was applied to the validation and test sets. The checkpoint with the lowest validation loss was evaluated on test sets. Having selected the best architecture based on ablation results, the final model BAT was trained on the full background augmented dataset from scratch using one NVIDIA H100 GPU with a batch size of 128 for 200 epochs.

\section{Results}\label{sec:simulation}

\begin{figure}[!htb]
    \centering
    \begin{subfigure}[t]{0.49\linewidth}
        \centering
        \includegraphics[
            width=\linewidth,
        ]{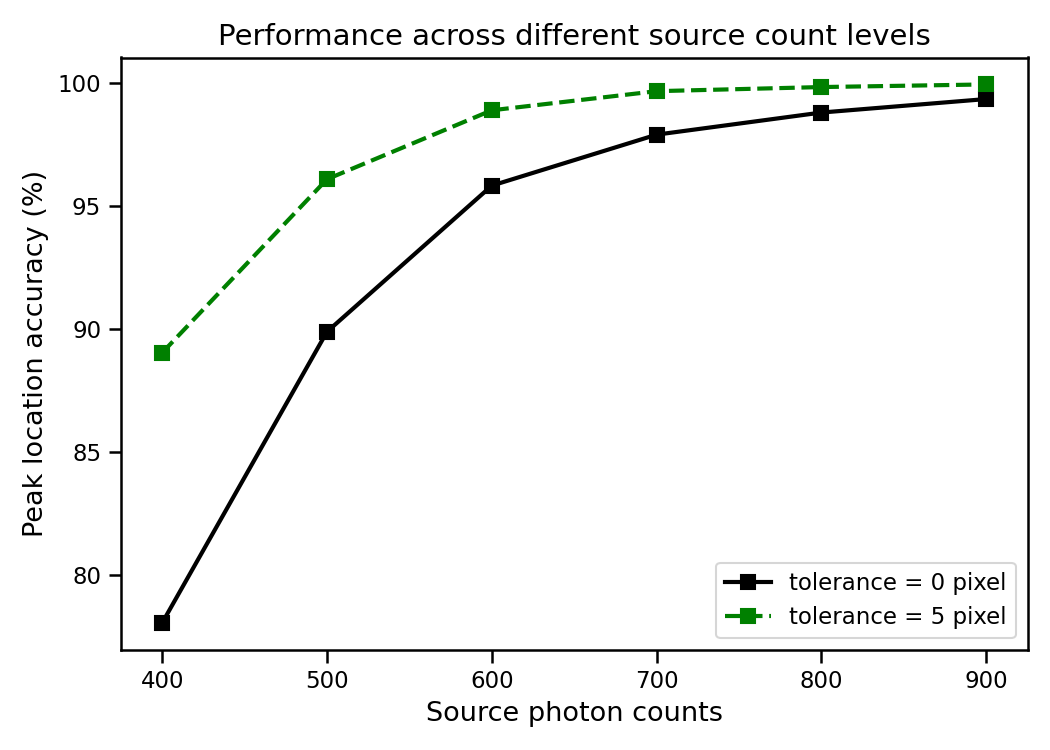}
        \caption{Trained on ideal detector data}
        \label{5a}
    \end{subfigure}
    \hfill
    \begin{subfigure}[t]{0.49\linewidth}
        \centering
        \includegraphics[
            width=\linewidth,
        ]{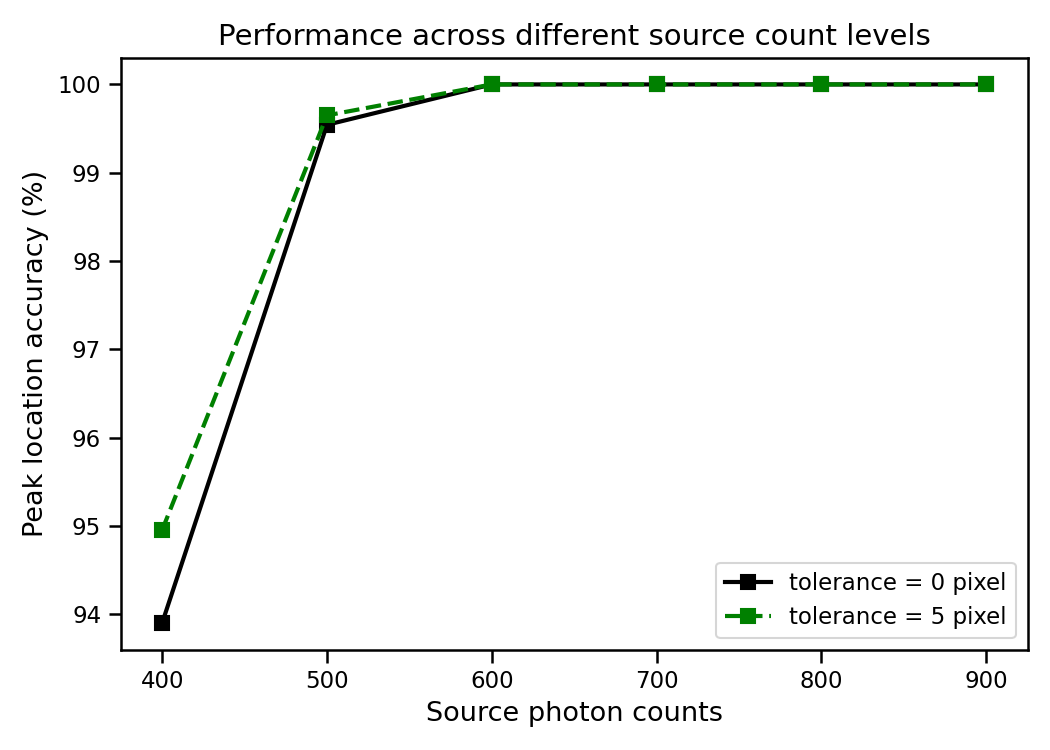}
        \caption{After background training}
        \label{5b}
    \end{subfigure}
     \vspace{0.3cm}
    \caption{Model generalization at lower source count levels}   
\end{figure}

\begin{figure}[htb]
    \centering
    \includegraphics[width=0.49\linewidth]{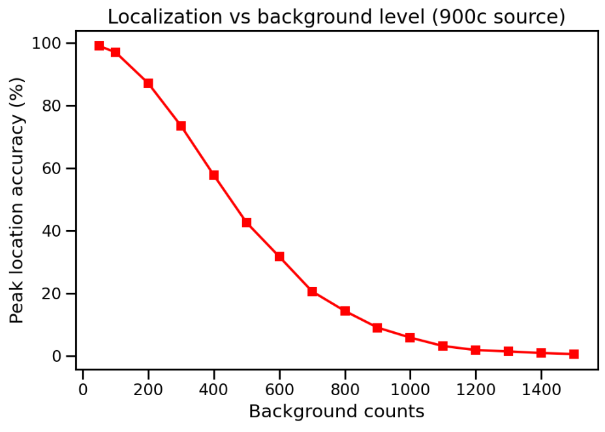}
    \includegraphics[width=0.49\linewidth]{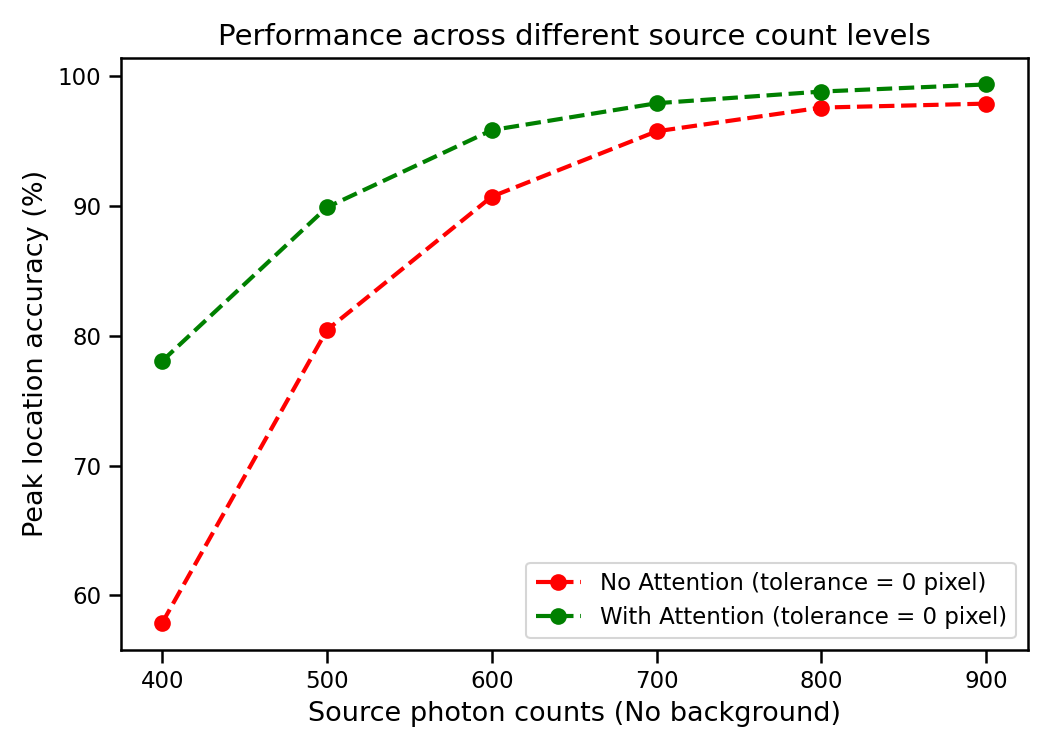}
    \caption{(Left) Performance degradation with increasing background in a model trained under ideal background conditions (Right) Lower count generalization degrades without attention.}
    \label{6}
\end{figure}

The primary objective is to predict whether an event has occurred and to accurately locate its coordinates. We report performance under two localization criteria: an exact-pixel match between the predicted peak and the true source location, and a criterion allowing a positional error of up to five pixels. Localization accuracy is the fraction of test samples that satisfy each criterion. Figure-4 shows an example in which the source position is reconstructed from a test detector measurement.
\begin{figure}[htbp]
    \centering
    \includegraphics[width=1\linewidth]{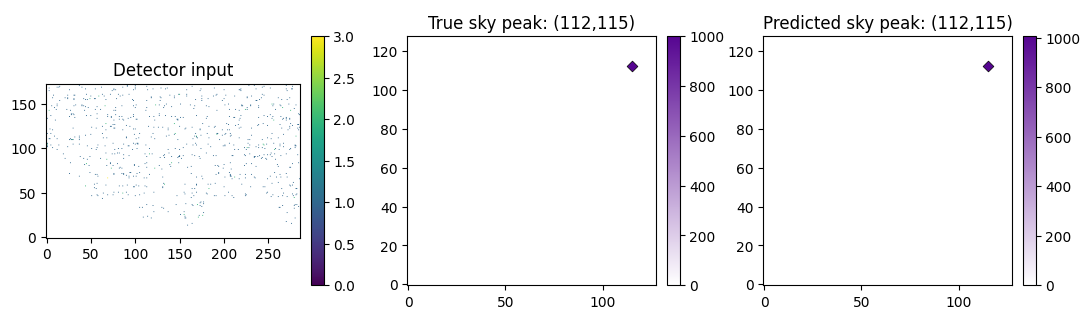}
    \vspace{0.3cm}
    \caption{Sample Reconstruction}
\end{figure}
Figure~\ref{5a} shows the performance of the model trained on ideal background with only 900 source counts. The model generalizes below its training level, achieving over 95\% exact-pixel localization accuracy at 600 counts but falls to approximately 78\% at 400 counts. Testing on background corrupted samples reveal that out of distribution background causes a rapid drop in localization accuracy as shown in Figure~\ref{6}. Performance remains above 80\% for only up to approximately 200 background counts, beyond which the model increasingly fails to localize sources.
The final model, trained on the background-augmented dataset, performs substantially better at predicting source locations at low detector SNR levels. At 900 source counts, it maintains at least 95\% exact-pixel localization accuracy with up to 7{,}500 added background counts, followed by a gradual decline at higher background levels. Although training remained fixed at 900 source counts, generalization to lower counts also improved under background-free conditions, with exact-pixel localization accuracy exceeding 99\% across the tested range of 500–900 source counts (Figure~\ref{5b}).

\begin{figure}[htbp]
    \centering
    \includegraphics[width=0.9\linewidth]{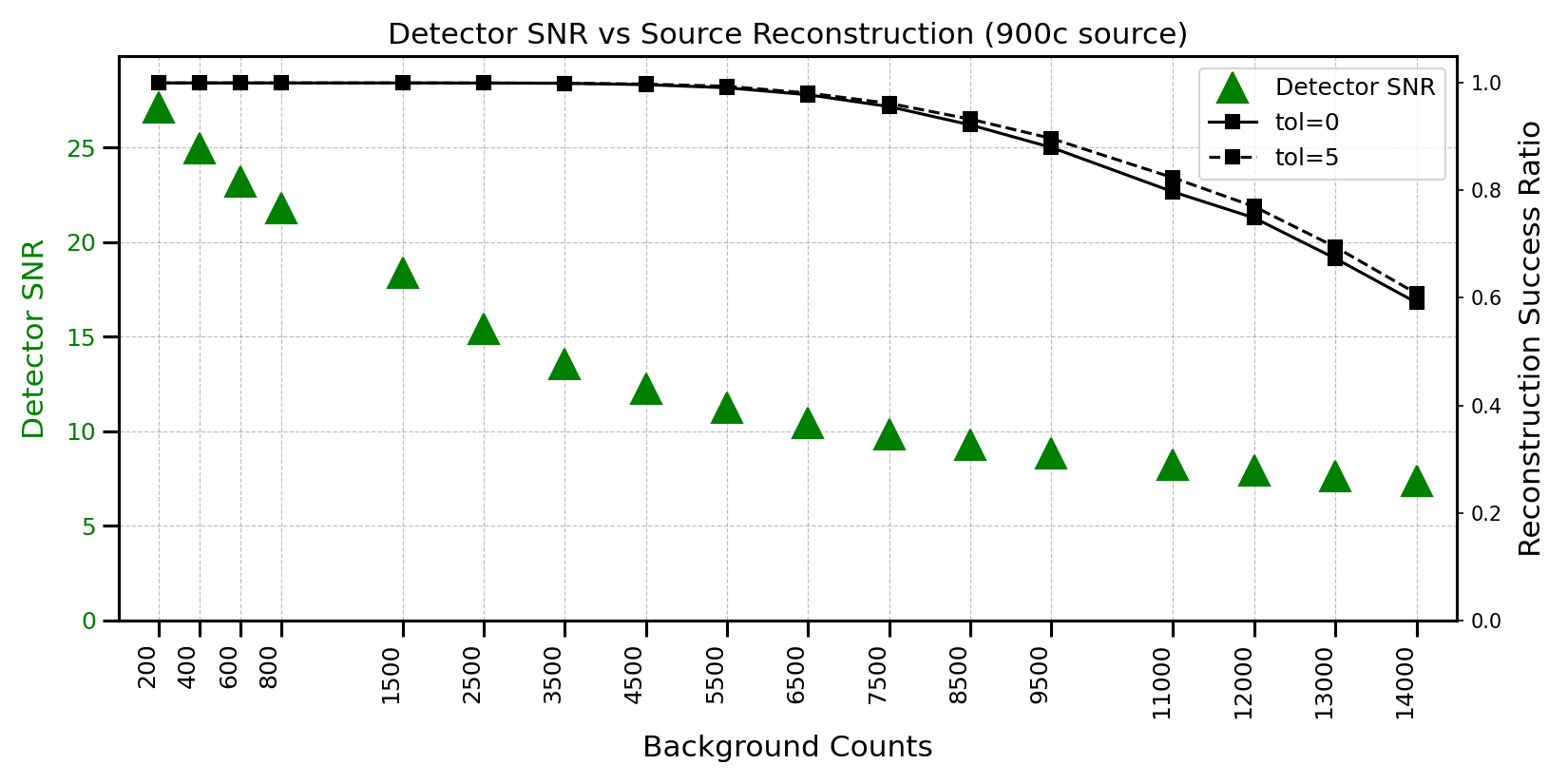}
    \vspace{0.3cm}
    \caption{Performance at lowering detector SNR conditions after background training}
    \label{9}
    \end{figure}

The BAT instrument operates in an environment of fluctuating background that scales the signal to noise ratio with the square root of detector counts. The detector plane SNR can be estimated by
\begin{equation}
    \mathrm{SNR} = \frac{N_s}{\sqrt{N_s + N_b}},
    \label{eq:snr}
\end{equation}
where $N_s$ is the number of source counts and $N_b$ is the number of background counts. Note that on the detector, $N_b$ and $N_s$ are not independently distinguishable. For on-board source localization, a minimum SNR of 6.5$\sigma$ is typically required \cite{threshold}. Figure~\ref{9} shows that our current best model has a success rate of 67.3\% in localizing sources with an expected SNR of 7.6$\sigma$ which corresponds to 13{,}000 background counts and lies beyond the 10{,}000 count maximum used in training.

\begin{figure}[htbp]
    \centering
    \includegraphics[width=0.48\linewidth]{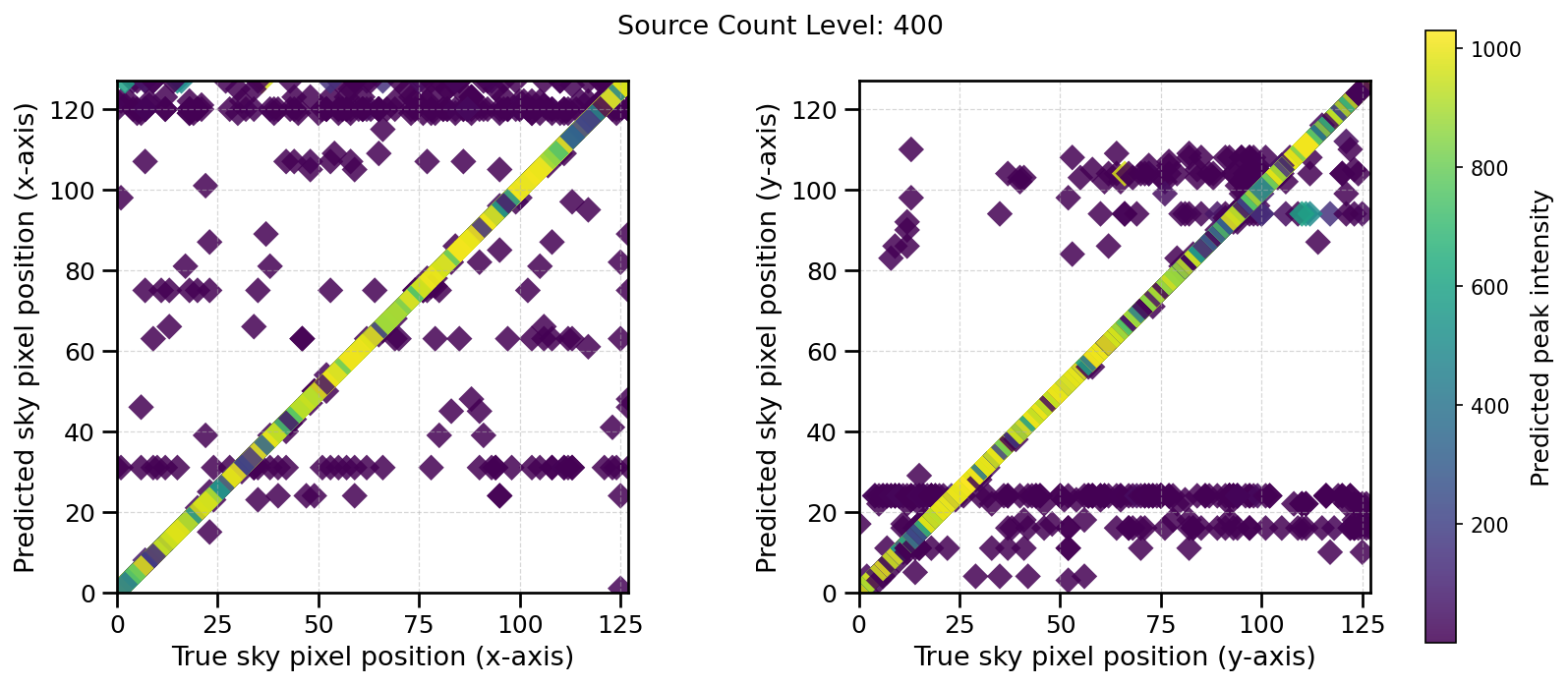}
    \hfill
    \includegraphics[width=0.48\linewidth]{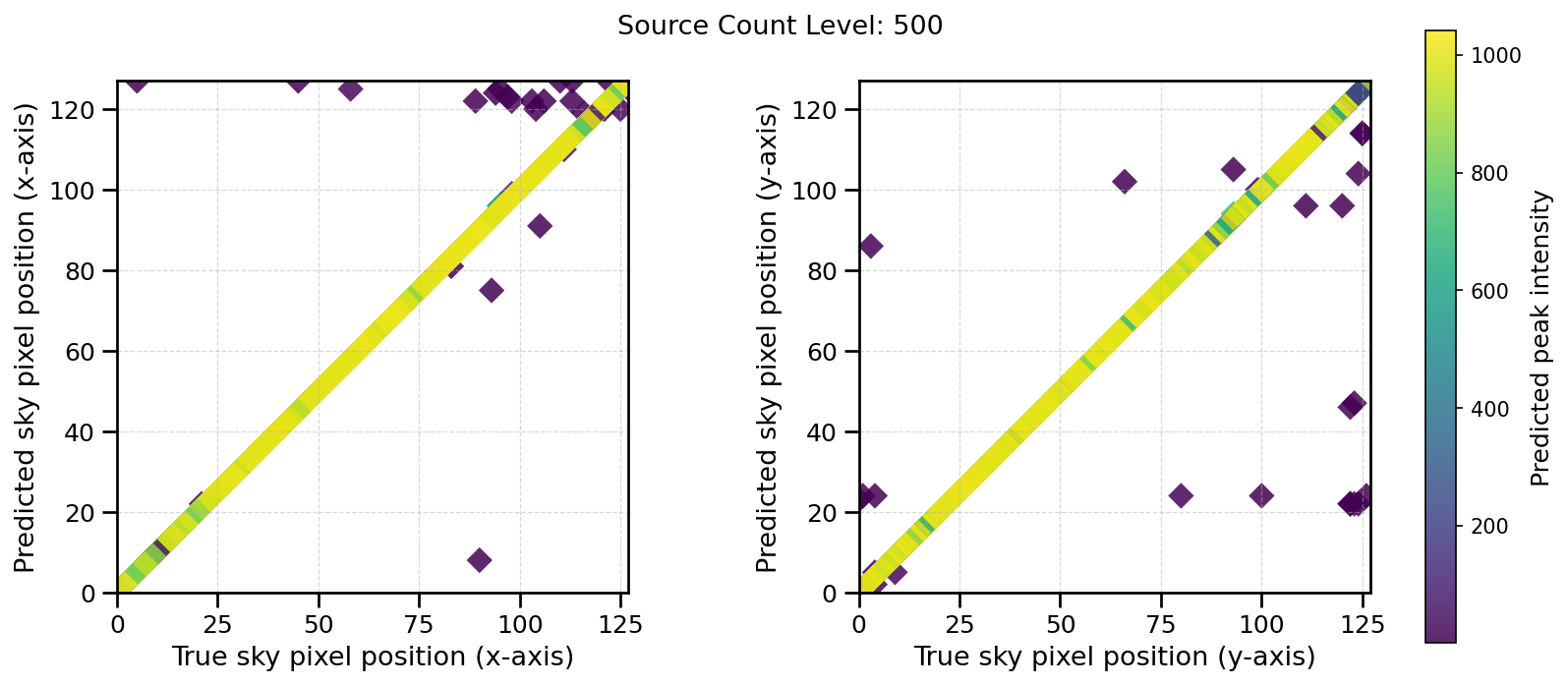}
    \hfill

\includegraphics[width=0.48\linewidth]{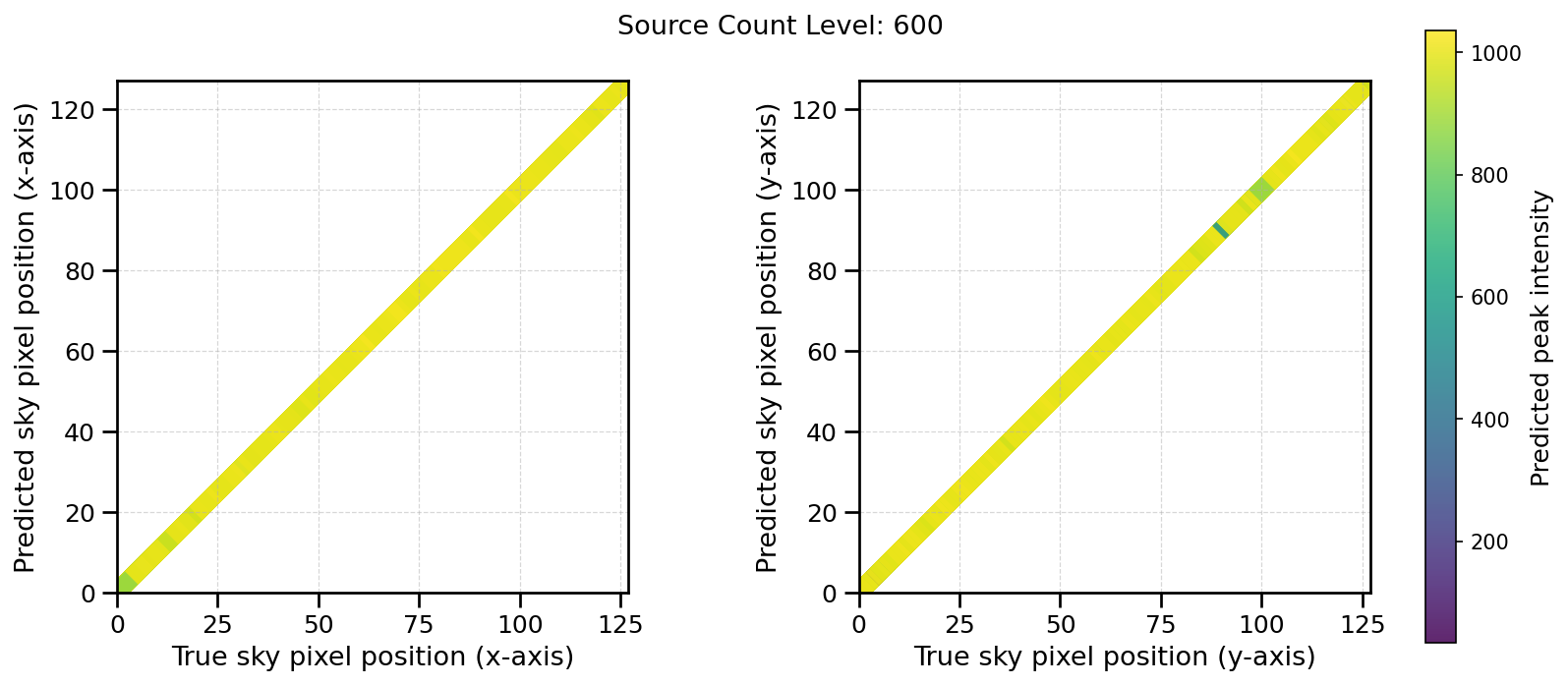}
    \hfill
    \includegraphics[width=0.48\linewidth]{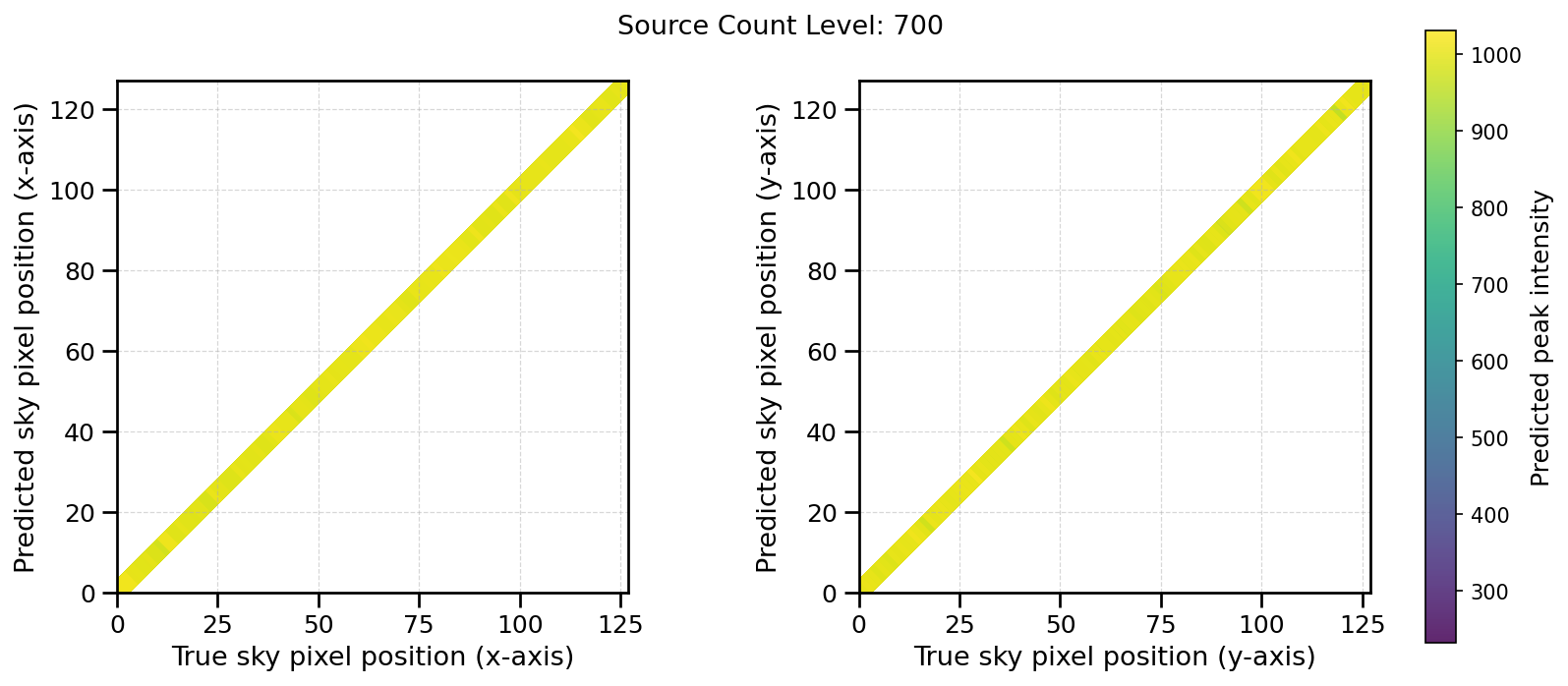}
\hfill

\includegraphics[width=0.48\linewidth]{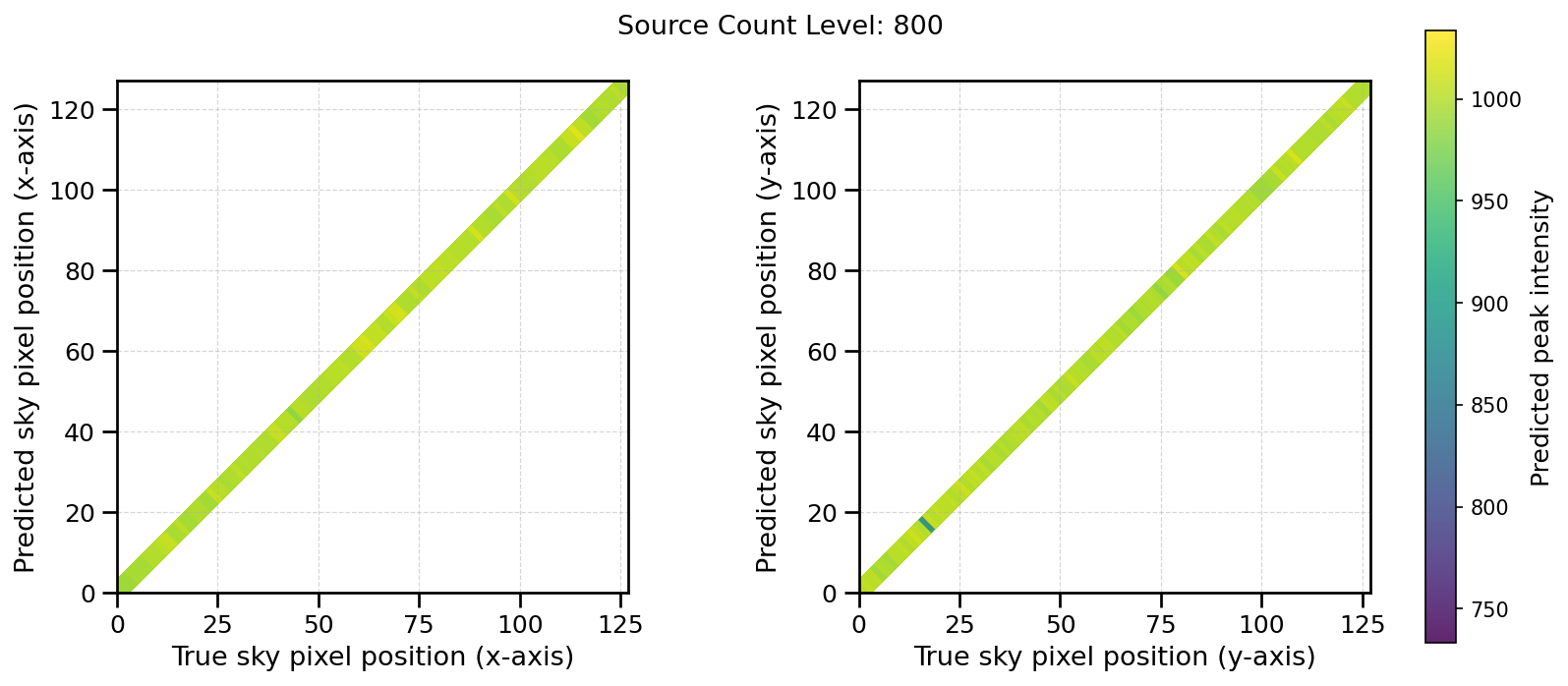}
    \hfill
    \includegraphics[width=0.48\linewidth]{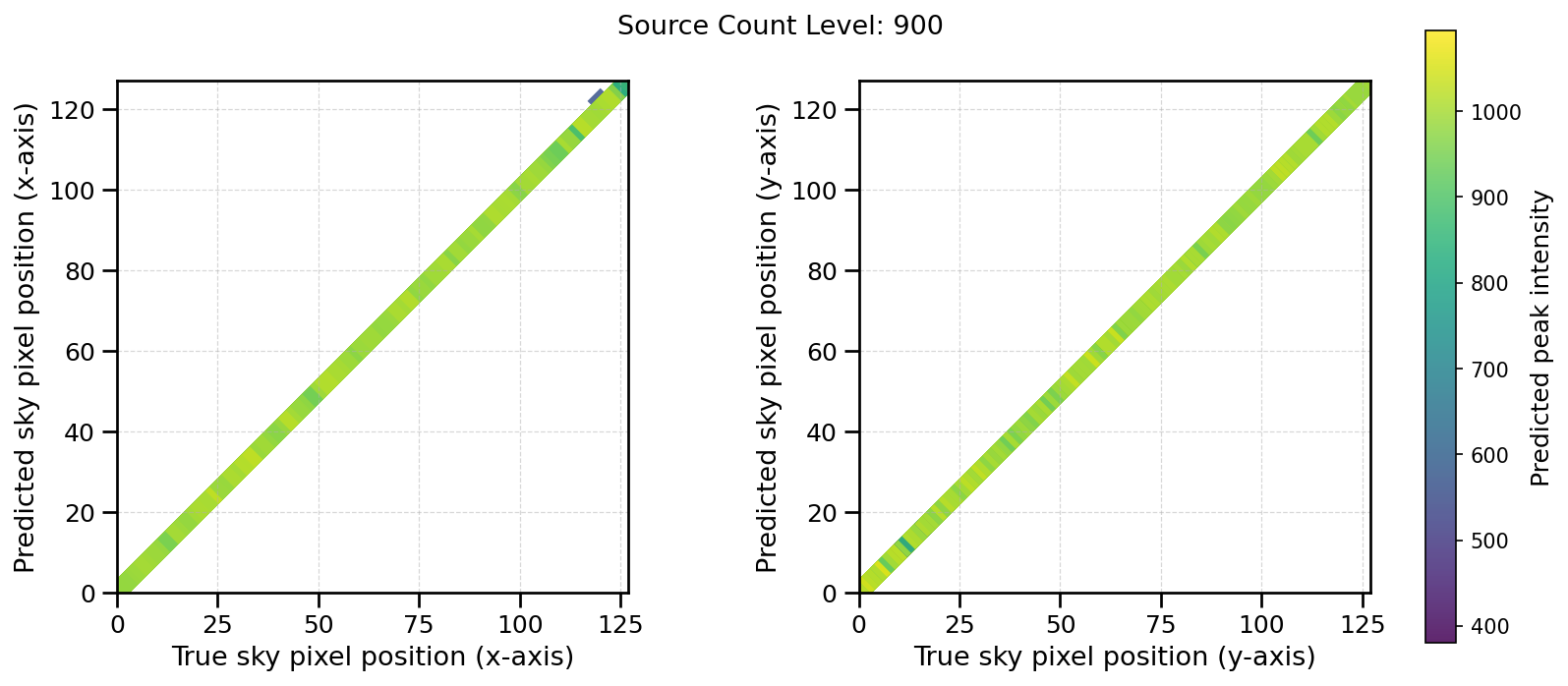}
    \hfill
    \vspace{0.3cm}
\caption{Positional error at lower source counts}
\label{fig-7}
\end{figure}
\begin{figure}[htbp]
    \centering
    \includegraphics[width=0.48\linewidth]{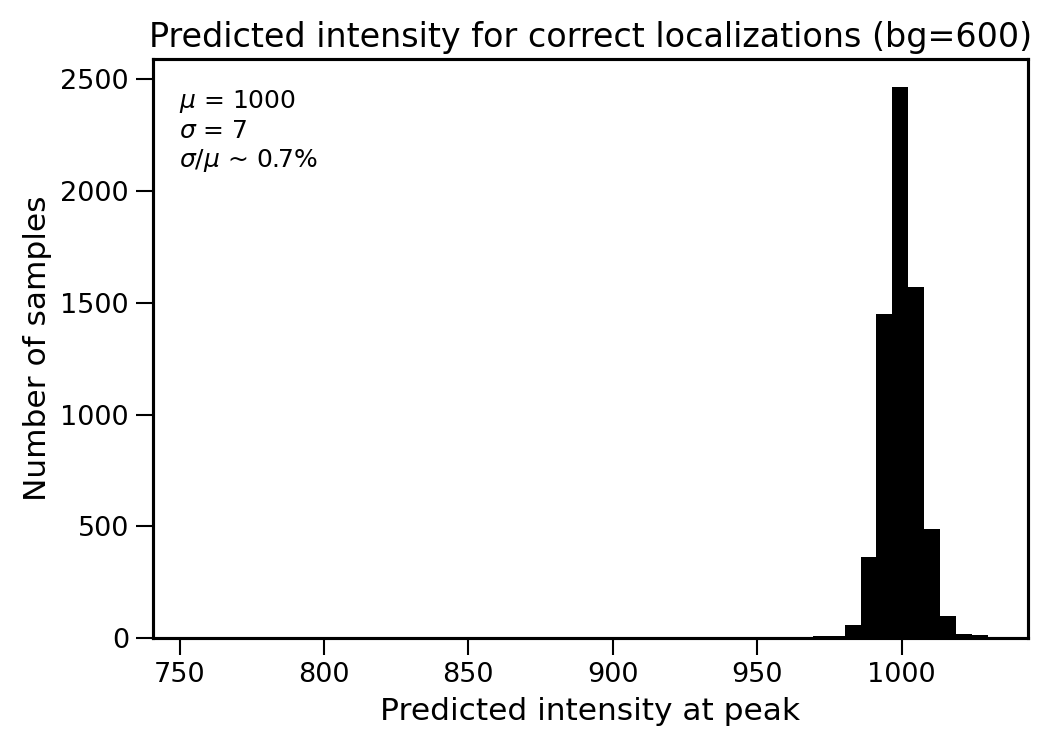}
    \hfill
\includegraphics[width=0.48\linewidth]{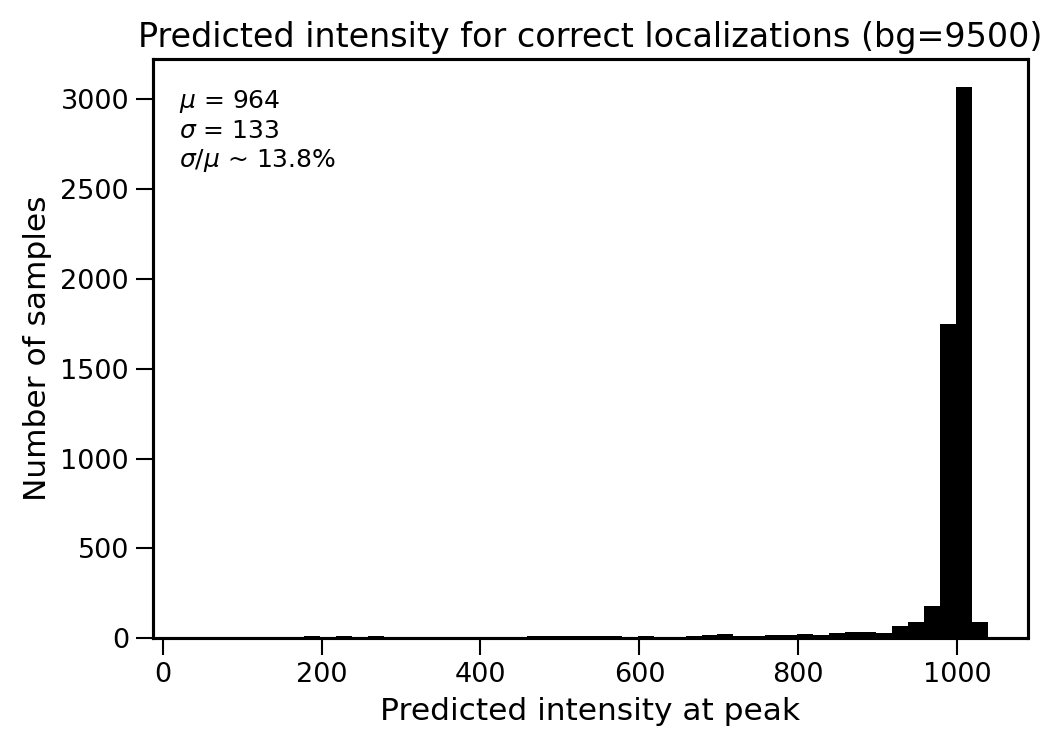}
    \vspace{0.3cm}
    \caption{Intensity distribution of prediction under varying background counts for a 900 count source}
    \label{8}
\end{figure}

Removing transformer from the proposed model yields an accuracy of 97.88\% on 900 source count test set, only 1.48\% below the proposed model. However, the difference becomes significant in lower count test sets, as shown in Figure~\ref{6}. The accuracy gap widens to 20.18\% at 400 counts, indicating that attention contributes more strongly to generalization under increasingly sparse photon conditions. 
Figure~\ref{fig-7} demonstrates how recovered intensity reflects the rise in uncertainty with lowering source counts. Correctly localized sources with near zero positional error tends to get assigned a high peak intensity close to the true value of 1000. More importantly, when a prediction is inaccurate, the model is successful in suppressing reconstructed intensity across all the levels tested. So the predicted peak intensity may serve as an implicit confidence property of the model that is useful in analyzing measurement at uncertain conditions. In Figure~\ref{8}, the fractional intensity scatter $\sigma/\mu $ is measured to be around $0.7\%$ at a relatively low background level of 600 counts. This spread widens to approximately $13.8\%$ at a high background level of 9{,}500 counts, reflecting the increased uncertain conditions at the detector.

\FloatBarrier
\section{Conclusion}\label{ref:conclusion}
This work establishes the feasibility of learning a reconstruction from simulated responses of a characterized planar system under low photon conditions. The results show that source positions remain recoverable well below the training level of 900 source counts. Under background-free conditions, the model achieved over 93\% exact-pixel localization accuracy with as low as 400 detected source photons. The background experiments show why performance on ideal measurements alone is insufficient for realistic systems. Training with background substantially improved robustness to lower detector SNR while also improving lower source count generalization. Incorrect localizations tend to have lower reconstructed peak intensities, but some correctly localized sources at low counts also have weak peaks. Peak intensity may therefore help identify uncertain estimates, although its usefulness as a confidence measure remains to be tested. Future work will extend training and evaluation across the full field, and test its applicability to build systems with improved off-axis sensitivity.

\section*{ACKNOWLEDGMENTS}
Shayla Sharmin and Branden Allen are supported by a special appropriation of the State of Hawai’i for the creation of the Space Science and Engineering Initiative (SSEI).
The University of Hawai'i's Koa high performance computing cluster was used in this work; the technical support and advanced computing resources from University of Hawai`i Information Technology Services –- Research Cyberinfrastructure, funded in part by the National Science Foundation CC awards 2201428 and 2232862, are gratefully acknowledged.

\bibliographystyle{spiebib}
\bibliography{report} 

\end{document}